\documentclass[prl,aps, twocolumn, showpacs]{revtex4}
\begin{document}

\title{Reply to Comment on ``Global positioning system test of the local position invariance of Planck's constant"}
\author{J. Kentosh}
\email{james.kentosh.336@my.csun.edu}
\author{M. Mohageg}
\email{makan.mohageg@gmail.com}

\affiliation{Department of Physics and Astronomy, California State University, Northridge, Northridge, California 91330-8268, USA}
\date{\today}

\begin{abstract}
We reply to a Comment [Berengut and Flambaum, Phys. Rev. Lett. {\bf 109}, 068901 (2012),  arXiv:1203.5592] on our recent Letter [Kentosh and Mohageg, Phys. Rev. Lett. {\bf 108}, 110801 (2012),  arXiv:1203.0102]. The Comment objects to our work partly because $h$ has dimensions and therefore measuring any variation of it is meaningless.  We reply that the relevance of dimensions in the study of fundamental constants has been the subject of some debate.  We also note that our model differs from the one normally used to compare hydrogen masers and cryogenic optical resonators. Ours is based on a different treatment of proper length and energy to conform to GR.  We also enclose in an Appendix our transmittal to the editors of Physical Review Letters.
\end{abstract}


\maketitle

$\pagebreak$

$\pagebreak$

{\bf Kentosh and Mohageg Reply:}  We do not agree that it is meaningless to measure variations of constants with dimensions.  Systems of units can be devised to set Planck's constant $h=1$ or $h=6.626 \times 10^{-34}$ J-s, but in any case the invariance of values or units assigned to $h$ must be verified through experiment.  There has been debate on this topic (e.g., \cite{duff02, duff04}), which cannot be resolved here.  Nevertheless, we acknowledged in our Letter \cite{kentosh} that to prove $h$ invariant in the context of the standard model extension is probably not possible with existent experimental evidence.  That is why we were careful to condition our findings as being applicable only within the context of general relativity (GR).

General relativity requires the invariance of proper mass, proper length, and the speed of light c.  The assumed invariance of those macroscopic variables provides the starting point for our work.  Rather than an absolute measure of $h$, our work provides a consistency check of GR based on those assumptions.

We also understand the importance of dimensionless ratios in experiments.  Most of our analysis focused on calculating limits for the dimensionless parameter $\beta_f$, which represents how closely atomic clocks on GPS satellites track the prediction of relativity based on the observable dimensionless ratio $f_{xo}/f_o$. That analysis compared satellite clocks to ground-based clocks.

The Comment \cite{berengut12} is incorrect in implying that we have relied on a single type of clock to measure variations in $h$.  The Letter describes our use of clock comparisons between cryogenic optical resonators (CSOs) and hydrogen masers.  That data provides a second type of clock, used to estimate $\beta_t$, which represents how closely CSOs would track relativity in a GPS orbit.

In response to the Comment we further explain how we extracted information on $\beta_t$ and $h$ from our estimate of $\beta_f$:  The invariance of $c$ is a necessary part of relativity.  The invariance of proper mass is required by the equivalence principle.  Once mass and $c$ are defined, energy is also known, not only for massive bodies but for their constituent particles and emissions.  We used this to support the invariance of the proper energy of atomic transitions.  Thus, the rate of an atomic clock depends on the local value of $h$.

The invariance of proper length in GR is less obvious.  Relativity is based on the invariance of the line element $ds$, which can be related to proper length.  Such invariance would be meaningless if the lengths of material bodies were to differ from length predicted from $ds$.  The concept of rigid measuring rods is essential to relativity.  Therefore, we conclude that in GR the proper lengths of all bodies should be invariant.

We believe that the invariance of length, energy and $c$ satisfies the insistence in the Comment that it is not meaningful to discuss limits on $h$ ``unless it is made clear what units are arbitrarily being held fixed."

$\pagebreak$

We used the invariance of proper length to deduce how a CSO would function within the framework of GR.  The resonant frequency of a CSO is determined by the round trip transit time for a wave within the cavity.
If the proper length of a CSO's resonating cavity is invariant, then its proper clock rate depends on $c$, unlike an atomic clock.  With a constant $c$ its rate is invariant.  This interpretation should satisfy the suggestion in the Comment that when comparing two clocks, ``some theoretical calculation is required to interpret the result in terms of LPI violation."

Finally, we note that the equation in the Comment, $\beta_{\text{H-maser}} - \beta_{\text{CSO}} = 3 \kappa_{\alpha}+\kappa_e - 0.1 \kappa_q$, relies partly on a simple heuristic model \cite{turneaure} and that other models are possible.  Ours is based on a different treatment of proper length and energy.

If the fine-structure constant $\alpha$ were found to violate LPI, terrestrial clock comparisons, by themselves, would be unable to determine which of the variables comprising $\alpha$ also varies.  Different experimental approaches such as ours are needed to answer that question.


\pagebreak

\begin{appendix}
\section{Appendix \--- Transmittal of the draft Reply to the editors of Physical Review Letters}
We appreciate the interest of J. C. Berengut and V. V. Flambaum (B\&F) in our work.  They raise issues that are important for the study of local position invariance (LPI) and general relativity (GR).  We have received similar comments from others (e.g., [A]) and welcome the opportunity to respond.  We have submitted a formal Reply for consideration by Physical Review Letters.  This transmittal provides our supplemental comments, which are not intended for publication.

There is an interesting story to be told here.  B\&F, and many others, adhere to a paradigm that claims that it is meaningless to measure (or even to discuss) changes in fundamental constants that have units.  Our Letter attempts to avoid that obstacle by using general relativity as a basis for fundamental units.  Have we been successful?  The core issues are worthy of a thoughtful discussion.  Our version of the story is partly described above and in our Letter;  there is much more that can be said.  Publishing the Comment and our Reply would provide an opportunity to expound on our interpretations.  Perhaps that would be of interest to the readership of PRL.

Nevertheless, we are concerned about some of the wording in the Comment that mischaracterizes our work.  The Comment strongly implies that we based our findings on a single type of clock.  A careful reading of our Letter would show that we did not rely solely on atomic clocks, whatever one might say of our approach.  There is nothing in the Comment that acknowledges our attempt to work within the framework of GR and to limit our results thereby.  At the least, we suggest that the Comment be revised to more fairly represent our work.  Of course if it is changed substantially, we should be allowed to revise our Reply accordingly.

On a minor point, B\&F's discussion of our Eq. (6), immediately following Eq. (1) in the Comment, incorrectly concludes that specification of units is required.  In fact, our Eq. (6) contains the ratios of variables with the same units and is dimensionless.  If B\&F mean to say that $h$ itself has units, then those units originate from the definition of $h$ and not from our equation.

As for the equations cited by B\&F, some of the work cited in the Comment relies on an implicit assumption that $h$ is invariant.  The most obvious example is found in their Ref. [3].  Eq. (1) in Ref. [3] describes the energy of an atomic level.  The variational of that expression is taken in Eq. (2) of Ref. [3], where it is clear that $h$ has been treated as invariant.  It is puzzling that a study of the possible variation of fundamental constants begins with an assumption that one of the most important constants in physics is invariant.

Equally puzzling is the treatment of length in Appendix A of our Ref. [5] (of the Reply), the origin of the $3 \kappa_{ \alpha}$ term.  The frequency of a CSO is presumed to change with $\alpha$, which causes a change in wavelength.  But it is not clear how that works from the perspective of proper length.  Presumably, in the local reference frame of the resonator, the frequency should be invariant, as with our approach.  If their frequency is written from the perspective of a distant reference frame, then they should account for the different time dilation there as we have done in our Letter.  Ref. [5] describes its approach as a simple heuristic model, and we believe our model is equally valid in the context of GR.

It is outside the scope of our Letter and Reply to discuss in detail the equations in the Comment.  In principle our two approaches should be compatible.  We had originally thought that theirs would reduce to ours in the limit of GR.  B\&F have not pointed out any specific errors in our approach.  It is clear that they have solved the problem their way and we have solved it our way.  The final answer could be something altogether different.

As for B\&F's belief that measuring variations of dimensional constants is meaningless, there is much to be debated there, but this is not the time or place.  We merely note that in the derivation of their equation (Appendix A of Ref. [5] in the Reply), the expression for $\alpha$ is substituted for $h$ wherever it occurs.  They are simply expressing variations of $h$ in terms of $\alpha$.

We hope that our remarks are clear and reasonable, and we thank you for the opportunity to respond.

\begin{flushleft}
[A]  Jacob Bekenstein, Pers. Comm., March 4, 2012.
\end{flushleft}

\end{appendix}

\end{document}